\documentclass[]{spie}  
\usepackage[]{graphicx}

\title{Simulations of Noise in Disordered Systems}

\author{C. J. Olson Reichhardt\supit{a} and C. Reichhardt\supit{b}
\skiplinehalf
\supit{a}T-12, Los Alamos National Laboratory, Los Alamos, NM, USA \\
\supit{b}T-CNLS, Los Alamos National Laboratory, Los Alamos, NM, USA
}

\authorinfo{Further author information: \\C.J.O.R.: E-mail: cjrx@lanl.gov, Telephone: 1 505 665 1134\\  C.R.: E-mail: charlesr@cnls.lanl.gov, Telephone: 1 505 665 0059}

\begin{document} 
  \maketitle 

\begin{abstract}
We use particle dynamics simulations to probe the correlations between
noise and dynamics in a variety of disordered systems, including
superconducting vortices, 2D electron liquid crystals, colloids, 
domain walls, and granular media.  The noise measurements offer an
experimentally accessible link to the microscopic dynamics, such
as plastic versus elastic flow during transport, and can provide
a signature of dynamical reordering transitions in the system.
We consider broad and narrow band noise in transport systems, as
well as the fluctuations of dislocation density in a system near
the melting transition.  
\end{abstract}



\section{INTRODUCTION}
\label{sect:intro}  

Noise measurements have found fruitful applications
as a tool to probe both equilibrium and nonequilibrium
transitions.  An open question is whether phase transitions of
the type observed in equilibrium systems can also occur in
nonequilibrium systems, or whether it is even meaningful to
refer to ``phases'' in the latter case.  A particularly interesting
issue is whether there are similarities between the melting transition
of a thermal system and the disordered or plastic 
flow that occurs in a system driven out of equilibrium over
quenched disorder.  In this paper, we explore some
suggestive similarities in the noise properties of equilibrium
and nonequilibrium systems, based on our simulation
studies \cite{river,voltnoise,wigner,vortexchain,heterocolloid,stripe}
of a wide range of physical systems.  We are particularly interested
in possible connections between the dynamical heterogeneities observed
above the melting transition in a thermal system, and those found in the 
plastic or disordered flow state of a driven system.

We begin by considering the noise produced during the melting of a
vortex system confined to one dimension.  We compare this to
the case of a colloid system melting in two dimensions, and find
that similar features appear in the noise.  Finally, we consider
a driven system in the presence of quenched disorder, and show that as the
driving force is decreased, similar noise signatures appear as when
the temperature is increased in the thermal systems.  We identify
the heterogeneities responsible for the noise in each case, and
suggest how noise signatures can be used to further probe these
systems and increase our understanding of both melting and
plasticity.

Since this paper does not focus on a single system, but refers to work
performed on several different physical systems, we offer some very
brief background information here.
Ideal superconductors carry current without
resistance and perfectly expel externally applied magnetic
fields.  The superconductivity is destroyed when too great
a magnetic field is applied, and the material becomes a
normal resistive conductor.  Type-II materials remain
superconducting in high magnetic fields by allowing the
magnetic flux to penetrate the material in the form of
discrete quantized vortices which repel each other and
interact with defects in the superconducting material.  The
superconducting material returns to its normal resistive
state only at the center of these vortices;  the remainder
of the material still carries a supercurrent.  The
vortices experience a Lorentz force from the flowing current
and move through the superconductor until they are trapped,
or pinned, at defect sites. 
Another system of repulsive, overdamped particles which shows
many of the same dynamical features as the vortex system is
colloids, which are micron-sized particles suspended in a
liquid solution.  Each colloidal particle carries an electrostatic
charge, leading to a repulsive interaction between the colloids 
with a range that can be adjusted by varying the salt concentration
of the solution.  The colloids can be confined to two dimensions
by means of laser trapping, and can interact with quenched disorder
which may take the form of optical traps or a structured surface.

\section{Noise of one-dimensional melted systems}

To explore noise signatures near a melting transition,
we begin with the simplest case of a system confined to a one-dimensional 
(1D) channel.  Such a system can be created artificially by, for example,
using an optical trap to confine colloids to a 1D line \cite{clemens1D},
or by nanofabricating channels in a superconducting material, which then
serve to confine superconducting vortices 
\cite{Pruymboom,Theunissen,Besseling,Anders,Anders2}.
Quasi-1D confinement of vortices can also arise naturally due to 
trapping by extended defects such as grain \cite{Hogg,Gurevich} and
twin boundaries \cite{Duran}.  
Additionally, when a tilted field is
applied to layered superconductors, vortex chains appear in which a portion
of the vortices align in a string with a smaller lattice spacing than
that of the bulk vortices \cite{Bolle}.  We choose to model the vortex
chain system as an example of 1D confined melting.

For all of the systems described in this paper, we perform numerical
simulations of overdamped particles in a two-dimensional (2D) sample
with periodic boundary conditions in the $x$ and $y$ directions.
For the vortex chain state, we model the vortices as $N_v$ repulsive particles
which interact with weak random quenched disorder and thermal noise.
We confine a portion of the vortices to move along a quasi-1D channel.  The
equation of motion for a vortex $i$ is
\begin{equation}
\eta {\bf v}={\bf f}_i=-\sum_{j\ne i}^{N_v}\nabla_i U(r_{ij})+{\bf f}_{i}^{p}+{\bf f}_{i}^{T}.
\end{equation}
Here $\eta=1$ is the damping constant and $U_v=-\ln(r)$ is the vortex 
interaction potential, treated as in Ref~\cite{Jensen}.
The pinning potential arises from $N_p$ randomly spaced 
attractive parabolic traps of
range $r_p$ and strength $f_p=1.0$.
The thermal noise ${\bf f}_{i}^{T}$ arises from random Langevin kicks
with $<f^{T}(t)>=0$ and 
$<f^{T}_i(t)f^{T}_j(t^{\prime})>=2\eta k_B T \delta_{ij} \delta(t-t^{\prime})$
We measure temperature in units of the temperature at which the bulk
vortices melt, $T_m$.  
Further details of the simulation can be found
in Ref~\cite{vortexchain}.

\begin{figure}
  \begin{center}
    \begin{tabular}{c}
      \includegraphics[width=0.8\textwidth]{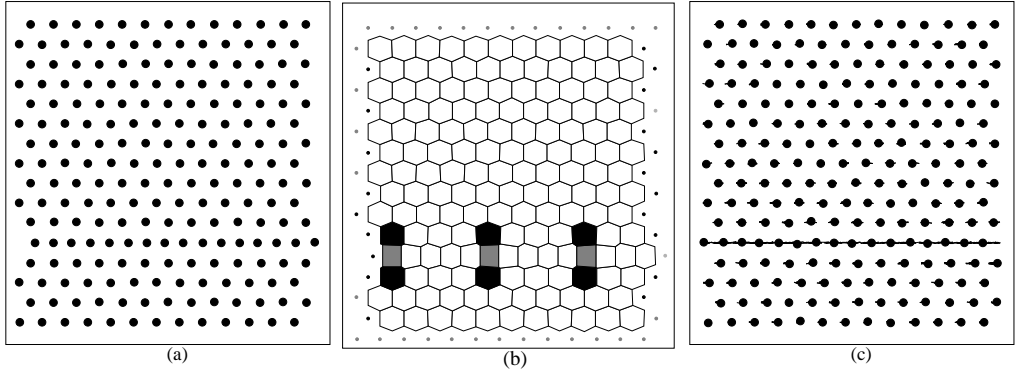}
    \end{tabular}
  \end{center}
  \caption[fig1]
  { \label{fig:chain1}
(a) Black dots: Vortices in the vortex chain sample.  The 1D channel of
increased vortex density is in the lower third of the panel. (b)
Voronoi construction indicating sixfold (white), sevenfold (black), and
fourfold (gray) coordinated vortices.  The three extra vortices inside
the channel each form a defect.  (c) Lines indicate vortex motion in
the system for a temperature $T/T_m^B=0.25$ above the transition to
motion along the chain.}
\end{figure}

To prepare the system, we place a triangular
array of vortices with lattice constant $a$ in the sample, and then put
additional vortices into the 1D confining channel, so that the lattice
constant inside the channel $a^{\prime}<a$, as illustrated
in Fig.~\ref{fig:chain1}(a).  In order to accommodate the incommensuration
between the vortices in the channel and the surrounding vortex lattice,
dislocations form along the channel.  These can be imaged by performing 
a Voronoi tessellation of the sample, as shown in Fig.~\ref{fig:chain1}(b).
At low temperatures, the incommensuration in the chain is stationary or
pinned in a well defined location.  As the temperature increases, the
vortices inside the chain become mobile at temperatures well below the
bulk melting transition $T_m^B$
of the triangular lattice, and begin to move back
and forth along the channel, as indicated in Fig.~\ref{fig:chain1}(c).
The transition from localized to mobile vortices inside the channel
can be regarded as a 1D melting transition.

\begin{figure}
  \begin{center}
    \begin{tabular}{c}
      \includegraphics[width=0.8\textwidth]{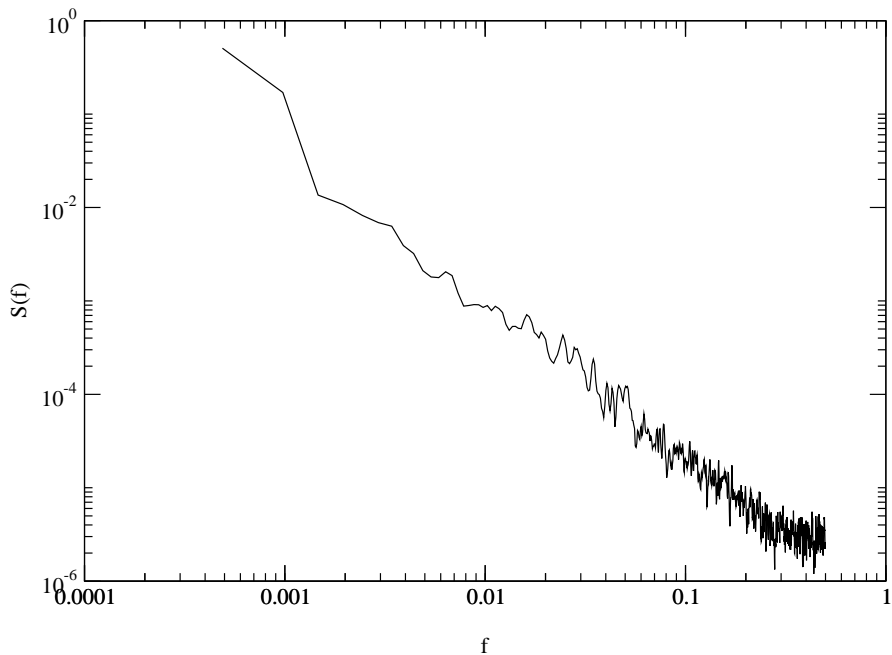}
    \end{tabular}
  \end{center}
  \caption[fig1]
  { \label{fig:chain2}
The power spectra $S(f)$ for the displacement of the vortex chain
$X$ at a temperature just above melting, $T/T_m^B=0.25$.  The spectra
has a $1/f^\alpha$ form, with $\alpha \sim 1.6$.  
}
\end{figure}

As a measure of the vortex motion along the chain at a given temperature,
we consider the net displacement $X$ of a vortex in the chain from
its starting position.  For temperatures just above where the vortex chain 
becomes mobile, individual vortices along the chain do {\it not} move 
continuously, but instead move in discrete jumps of nearly a lattice
constant in magnitude.  These jumps run through the chain as a soliton
like pulse with the vortices jumping sequentially.  The size of the jumps
is determined by the periodic potential created by the stationary
ordered bulk vortices.  The power spectra
$S(f)=|\int X(t)e^{-2\pi ift}dt|^2$ of the single vortex motion
just above melting has a $1/f^{\alpha}$ signature,
with $\alpha=1.6$, as shown in Fig.~\ref{fig:chain2}.  
This results from the discrete nature of the
jumps in $X$.  As the temperature further increases, the motion along
the chain becomes more continuous, and the magnitude of low frequency
power in $S(f)$ drops until, at high temperatures, the noise
spectra becomes white.  This indicates that a noise signature can be
used to probe how correlated the fluctuations in the system are
at temperatures close to and well above melting.

Note that the soliton-like pulse, as it moves through the sample, causes
an individual vortex to abruptly shift its position by approximately
a lattice constant.  Thus, in monitoring the position of one vortex
in the chain, we observe relatively long periods when the vortex is
essentially stationary, separated by very rapid motion to a new
position as the soliton sweeps past.  It may be more useful
to think of the soliton itself as the ``particle'' moving through
the system, since the soliton is diffusing continuously through the
chain.
The solitons can be directly identified by the 7-4-7 trios of dislocated
vortices seen in Fig.~\ref{fig:chain1}(b).  Note also that the solitons
or dislocations,
which are not constrained to remain fixed to a given vortex, can move
through the system much more rapidly than individual vortices can.

\section{Noise of two-dimensional melted systems}

We next turn to melting in a 2D system.  There is much previous work
in this area, and a considerable number of studies have focused on
dynamical heterogeneities which occur at temperatures above melting 
when certain regions of the sample
have a higher mobility than the rest of the sample
\cite{Kob,Donati,Hurley,Kegel,Weeks,Cui,Tang}.  
These heterogeneities are both temporal, such as when a group of
particles moves along a stringlike structure \cite{Glotzer},
and spatial, due to the fact that the 5-7 dislocation pairs in the
lattice form highly inhomogeneous distributions 
\cite{Somer,Somer2,Tang,Marcus,Quinn,Chiang,Juan}.

Since we have observed interesting noise signatures related to 
motion along a 1D chain in the previous section, it is interesting to
ask whether similar noise signatures appear for stringlike motion that
is not confined to 1D.  To investigate this question, we have performed
simulations of colloidal particles as a specific realization of overdamped
repulsive particles that form a triangular lattice in the absence of
temperature.  In this case, the equation of motion for the colloids is
again given by Eqn. (1), but without quenched disorder, ${\bf f}_{i}^{p}=0$,
and with a screened Coulomb interaction,
\begin{equation}
U(r_{ij})=\frac{Q^2}{r}\exp(-\kappa r).
\end{equation}
Here $Q$ is the charge on the colloid and $1/\kappa$ is the screening
length, which can be adjusted experimentally by varying the salt 
concentration of the solution.  We take $\kappa=2/a$, where $a$ is
the lattice constant.  In each simulation, we fix the temperature to
a value above the melting transition, 
wait for $10^6$ time steps for any transients to
disappear, and then begin to take data.
Further details of the simulation appear in Ref~\cite{heterocolloid}.

\begin{figure}
  \begin{center}
    \begin{tabular}{c}
      \includegraphics[width=0.8\textwidth]{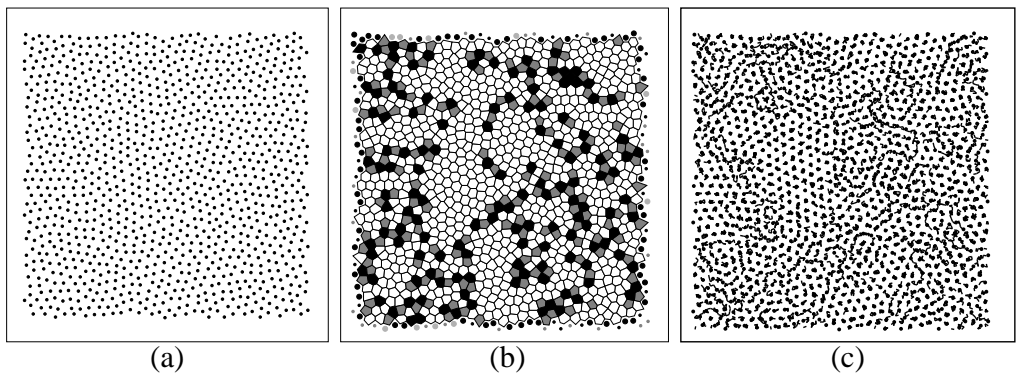}
    \end{tabular}
  \end{center}
  \caption[fig1]
  { \label{fig:hetero1}
(a) Black dots: Colloids in the two-dimensional melting 
sample at $T/T_m=1.04$, just above the melting transition.  (b)
Voronoi construction indicating sixfold (white), sevenfold (black), and
fivefold (dark gray) coordinated particles. 
(c) Lines indicate colloid motion over a fixed time interval.
}
\end{figure}

\begin{figure}
  \begin{center}
    \begin{tabular}{c}
      \includegraphics[width=0.8\textwidth]{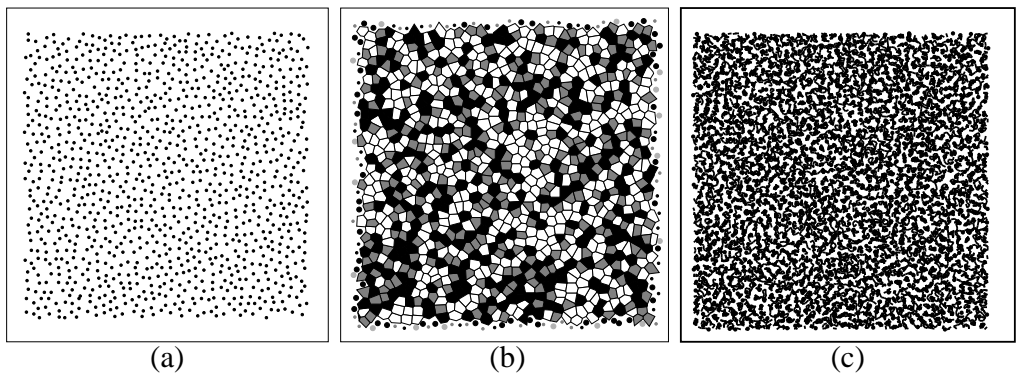}
    \end{tabular}
  \end{center}
  \caption[fig1]
  { \label{fig:hetero2}
(a) Black dots: Colloids in the two-dimensional melting 
sample at $T/T_m=7.0$, well above the melting transition.  (b)
Voronoi construction indicating sixfold (white), sevenfold (black), and
fivefold (dark gray) coordinated particles. 
(c) Lines indicate colloid motion over a fixed time interval.
}
\end{figure}

We illustrate the behavior of the system at two temperatures above
melting in Figs.~\ref{fig:hetero1} and ~\ref{fig:hetero2}.  Just above
melting, shown in Fig.~\ref{fig:hetero1}, about 30\% of the system
is filled with defects, but these defects are not evenly spread
throughout the sample.  Instead, the dislocations cluster into strings,
which form grain boundaries around sixfold-coordinated regions.
Particle motion is concentrated along the defect strings, as illustrated
in Fig.~\ref{fig:hetero1}(c).  This motion can be regarded as arising not
so much from the movement of the particles themselves, but rather as
being caused by the motion of the grain boundaries or defect strings.
In order for a defect string to move over by one lattice constant, the
particles currently inside the defect string must shift position in order
to join the ordered grain that is growing by one lattice constant as a
result of the motion of the boundary.  Thus, if we monitor the movement
of a single particle, we observe that the particle remains stationary
for a period of time before suddenly shifting position by a sizable
fraction of a lattice constant.  If we were, however, to instead monitor
the motion of a dislocation, we would find a more continuous motion
without long stationary pauses as the dislocation diffuses thermally.
It may in some cases be more useful to think of the dislocations themselves
as the primary ``particles'' whose dynamics are controlling the behavior
of the system, rather than the actual colloidal particles which we are
directly simulating.  The dislocations are more complicated entities than
the particles, however, both because they can be thermally created and
destroyed, unlike the colloids, and also because they do not remain fixed
to a single colloid but can move through the lattice much more rapidly
than an individual particle can.  There are clear analogies to the
1D case of diffusing dislocation discussed in Section 2, although
switching to two dimensions has introduced considerable complexity since
the lines along which the dislocations can move vary spatially.

When the temperature is increased to a value well above the melting
transition, the number of dislocations increases to fill
nearly half of the sample.  As illustrated in Fig.~\ref{fig:hetero2},
the motion of the colloids over time is now essentially homogeneous,
and the system is in a liquid-like state.  In this case, dislocations
are being created and destroyed so rapidly and easily that their
motions cannot be distinguished from those of the particles, which are
now moving freely through the sample.

\begin{figure}
  \begin{center}
    \begin{tabular}{c}
      \includegraphics[width=0.8\textwidth]{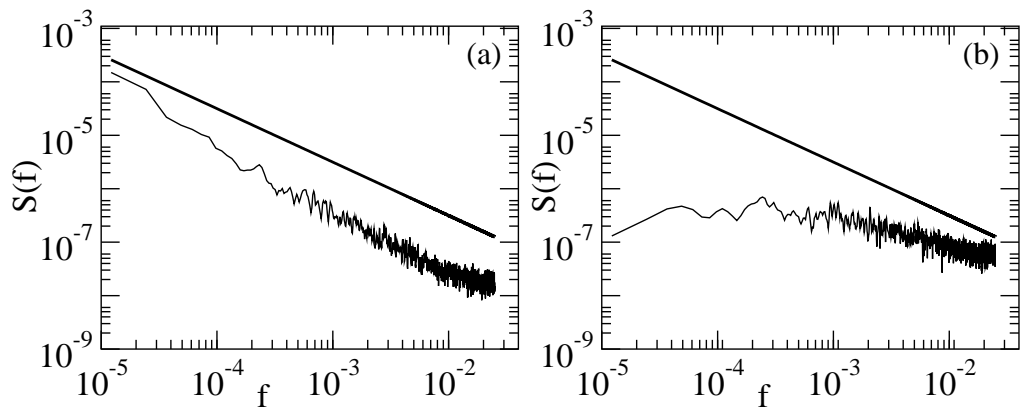}
    \end{tabular}
  \end{center}
  \caption[fig1]
  { \label{fig:hetero3}
(a) The power spectrum for the time series of the 
density of dislocations in the colloid system
just above melting at $T/T_m=1.04$.  The solid line indicates a
slope of $1/f$.  (b) The power spectrum for the time series
at $T/T_m=7.0$ along with a $1/f$ line.
}
\end{figure}

Since the motions of the dislocations play such a strong role in the
particle displacements just above melting, we use the dislocations 
themselves as a means of characterizing the behaviors of the system
at the different temperatures.  
We compute the defect configuration and density every
20 time steps for 20000 frames and obtain a time series of the defect
density for several temperatures.  
Just above melting we observe long
time fluctuations of the dislocation density, corresponding to the
persistence of the chain-like structures shown in Fig.~\ref{fig:hetero1}.
In contrast, at higher temperatures well above melting, the fluctuations
are very rapid.
In Fig.~\ref{fig:hetero3}(a) we plot the
power spectrum $S(f)$ of $P_6$ for $T/T_m=1.04$, which fits well to
a $1/f^{\alpha}$ scaling over more than three decades with the best
fit $\alpha=1.04$, close to $1/f$ noise.  As the temperature increases,
the spectrum changes from $1/f$ to white noise ($\alpha=0$) for low
frequencies, as shown in Fig.~\ref{fig:hetero3}(b) for
$T/T_m=7.0$.  

\begin{figure}
  \begin{center}
    \begin{tabular}{c}
      \includegraphics[width=0.8\textwidth]{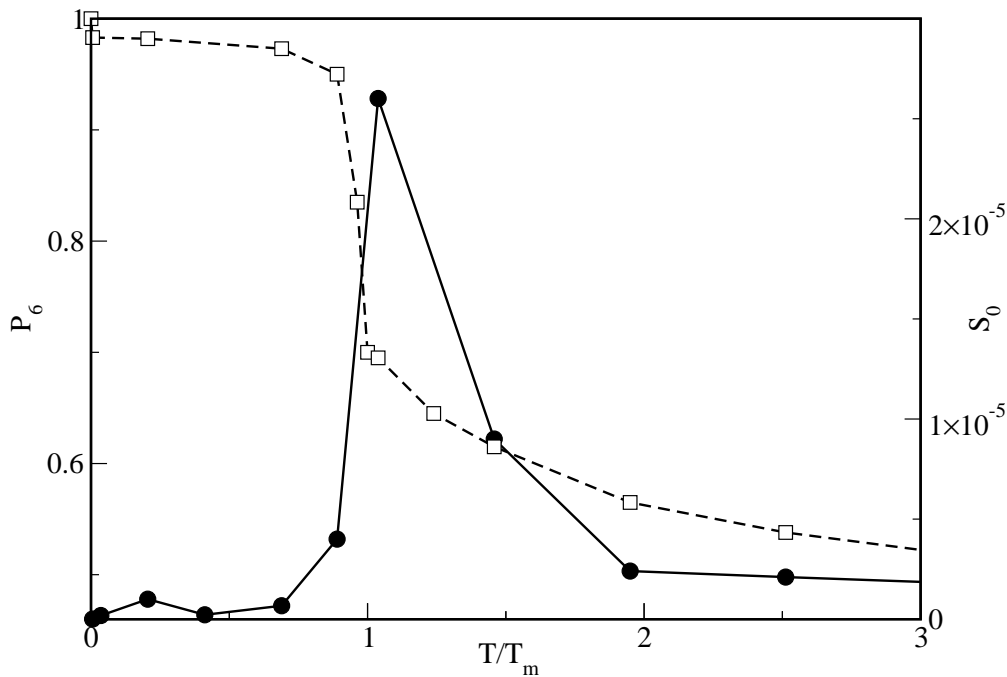}
    \end{tabular}
  \end{center}
  \caption[fig1]
  { \label{fig:hetero4}
Squares (dotted line): The fraction of sixfold coordinated particles $P_6$
versus temperature, with melting occurring at $T/T_m=1$.
Circles (solid line): The integrated noise power $S_0$, obtained from
$S(f)$, shows a peak just above the melting transition.
}
\end{figure}

It is clear from Fig.~\ref{fig:hetero3} that the amount of power at
low frequencies drops dramatically at the higher temperature, when
long-time fluctuations in the defect density are no longer present.
This suggests that the noise power could provide a useful means of
characterizing the system above melting.
We integrate $S(f)$ over the first octave of frequencies to obtain
the noise power $S_0$
\cite{Merithew,Rabin}, which we plot in Fig.~\ref{fig:hetero4}
as a function of temperature along with the defect density.
There is a prominent peak in $S_0$ which coincides with the
onset of the defect proliferation  above melting.
For increasing $T$, the noise power falls 
and saturates when the spectrum becomes white.  
Thus, just above melting, when the system is highly heterogeneous,
the noise power is maximized.  It is especially interesting that
$S_0$ appears to be diverging near $T/T_m=1$, suggesting that $S_0$ may
be a useful indicator of the melting transition.

\section{Noise of dynamically ``melted'' systems}

In this final section of the paper, we consider the similarities and
differences between noise in the thermodynamic systems discussed above
and noise in nonequilibrium systems.
There are a very wide range of systems falling into the category of
elastic media driven over periodic or random substrates, including
friction where atoms move over a periodic potential created by the
underlying stationary atoms \cite{Persson}, sliding charge-density
waves \cite{Gruner,Thorne},
and vortex motion in
superconductors with random quenched disorder \cite{Balents,LeDoussal}
or periodic hole or dot arrays 
\cite{Baert,Harada,Martin,VanLook,Reich2000,Surdeneau,LongPeriodic}

We add a term ${\bf f}_d=f_d {\hat x}$
to Eqn. (1) for the vortices in the 
1D vortex chain system described in Section 2.
This represents a uniform driving force applied to all of the vortices
in the system which, in the absence of quenched disorder, would cause
them to move in the positive $x$ direction, along the chain.
In an experiment, this force would arise when a current is applied
to the sample, creating a Lorentz force on the vortices that causes
them to move in the direction perpendicular to the applied current.
In the absence of temperature, $T=0$, as we slowly increase the
driving force $f_d$ we find that for low values of drive, $f_d<f_c$,
all of the vortices in the system remain trapped by the quenched
disorder and do not move.  At the critical force $f_c$, the first
onset of vortex motion occurs.  This motion can be detected experimentally
in the form of a voltage drop, since the normal core of each vortex
dissipates energy when the vortex moves.  The motion has also been observed
directly in imaging experiments \cite{Matsuda}.
In the case of the 1D
vortex chain system, the entire system does not depin simultaneously.
Instead, the vortices inside the 1D channel begin to move first, at 
a drive much lower than the drive at which the bulk vortices depin.
This is similar to the melting transition shown in Section 2, where
the vortices inside the channel begin to thermally diffuse at temperatures
that are much lower than the temperature at which the bulk vortices
begin diffusing.  The motion of the 1D chain of vortices is
the same as the motion shown in Fig.~\ref{fig:chain1}(c), with one important
difference.  In the thermally activated motion, the dislocations move 
diffusively to the left and right with equal likelihood, while in the
driven motion, the dislocations move only to the
right, giving a net transport of vortices in the direction of the
drive.  Just above the depinning force,
$f_d \sim f_c$, the motion occurs as a soliton pulse
\cite{LongPeriodic,Pruymboom,Theunissen,Besseling,Surdeneau}, 
with each vortex
jumping a single lattice constant and then remaining
stationary most of the time, as illustrated
in Fig.~3 of Ref~\cite{vortexchain}.  
As in the case of the melting system, the disturbance or incommensuration
moves much faster than the individual vortices.
As the drive is further increased,
however, more solitons appear and become more closely spaced, until
at higher drives the motion of the vortices in the chain becomes
continuous.  At these high drives the velocity response of the vortices
in the chain also becomes Ohmic and is linear with the drive.

Since there are similarities between the 1D system above melting and
above depinning, it is interesting to ask whether similarities also
occur for the 2D system.  There has been a large amount of work
exploring the dynamics of 2D driven vortex lattices 
interacting with disorder.  Experiments 
\cite{Bhattacharya,Yaron,Hellerqvist,Henderson,Duarte,Pardo,Marchevsky,Pardo98,Troyanovski}, 
simulations 
\cite{Koshelev,Brass,Shi,Dominguez96,Faleski,Spencer,Dominguez99,Moon,Ryu,voltnoise,Kolton},
and theory 
\cite{Koshelev,LeDoussal,Balents,Scheidl}
suggest that at low drives the vortex lattice
is disordered and exhibits plastic or random flow while at higher drives
the lattice can undergo a reordering transition and flow elastically.
In elastic flow,
each particle maintains the same nearest neighbors, whereas in plastic
flow, particles may move arbitrarily far away from their initial
nearest neighbors and the lattice tears.
In the highly driven state it was suggested by Koshelev and Vinokur
\cite{Koshelev}
that the flux lattice forms a moving crystal.  This is due to the fact
that, at high drives, the underlying quenched disorder begins to have an
effect similar to temperature on the lattice, with the effective temperature
decreasing as the lattice velocity increases.  Thus, the lattice can
freeze dynamically.
Such dynamical reordering has been observed in simulations
in vortex matter \cite{Moon,voltnoise,Kolton}, charge-density wave
systems \cite{Balents96}, and driven Wigner crystals \cite{wigner}.

\begin{figure}
  \begin{center}
    \begin{tabular}{c}
      \includegraphics[width=0.8\textwidth]{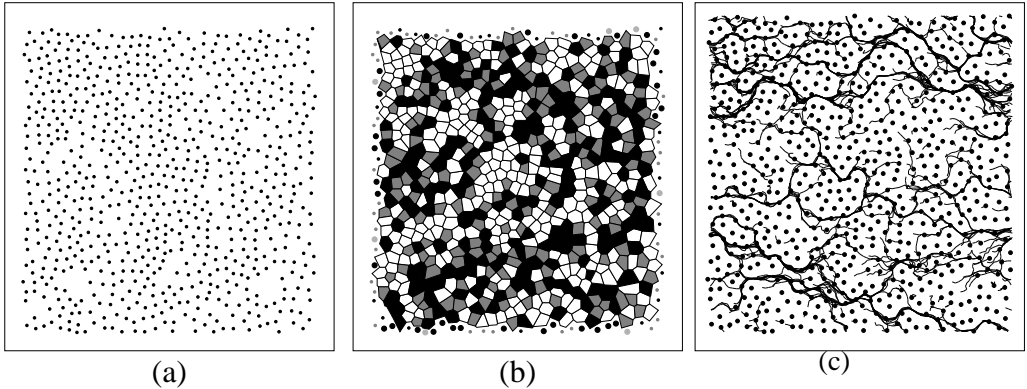}
    \end{tabular}
  \end{center}
  \caption[fig1]
  { \label{fig:colloid1}
(a) Black dots: Colloids in the two-dimensional driven sample 
in the plastic flow regime.  (b) Voronoi construction indicating sixfold
(white), sevenfold (black), and fivefold (dark gray) coordinated particles.
(c) Lines indicate colloid motion over a fixed time interval.
}
\end{figure}

\begin{figure}
  \begin{center}
    \begin{tabular}{c}
      \includegraphics[width=0.8\textwidth]{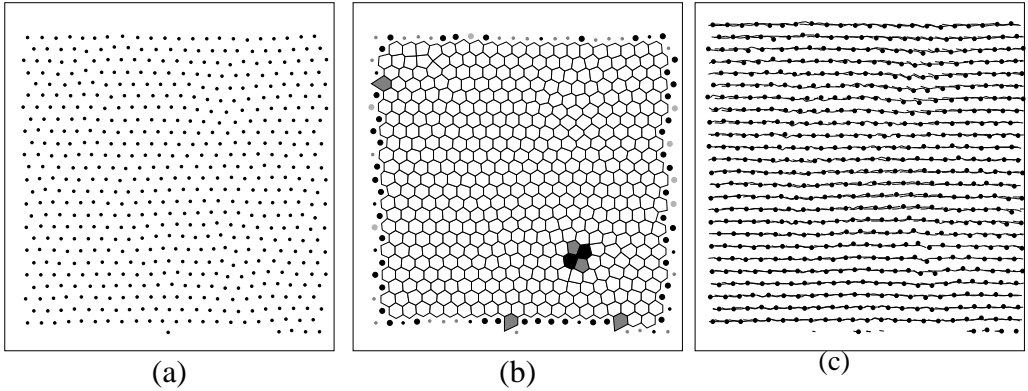}
    \end{tabular}
  \end{center}
  \caption[fig1]
  { \label{fig:colloid2}
(a) Black dots: Colloids in the two-dimensional driven sample 
in the elastic flow regime.  (b) Voronoi construction indicating sixfold
(white), sevenfold (black), and fivefold (dark gray) coordinated particles.
(c) Lines indicate colloid motion over a fixed time interval.
}
\end{figure}

We illustrate the motion of a 2D system of repulsive particles interacting
with a random substrate at two different values of the driving force
in Fig.~\ref{fig:colloid1} and Fig.~\ref{fig:colloid2}.
The system shown here is a colloidal sample with random pinning,
interacting according to the potential given in Section 3
(see also Ref~\cite{randomcolloid}); however,
other elastic media including vortices and localized charges have been
shown to produce the same behavior \cite{voltnoise,wigner}.
At drives just above depinning, the flow is plastic as indicated in
Fig.~\ref{fig:colloid1}.  Here the motion of the particles is clearly
heterogeneous, with only riverlike channels of movement occurring at
any given time.  
Similar channel structures, which resemble the fractal basins
created by natural rivers \cite{Rodriguez} 
have been observed in a large variety of
systems, including fluid flow in a disordered landscape, Josephson
junctions, Wigner crystals, magnetic bubbles, and stress networks in
granular systems.  
The heterogeneity shows certain similarities to the
heterogeneity just above melting in Section 3; however, here there is
a clear alignment of the paths of particle motion with the drive, 
providing a net transport of particles.  It is interesting to note that,
in the plastic flow regime, individual particles do not move continuously.
Instead, a particle will generally move only about a lattice constant before
displacing a particle from a pin and becoming trapped in the now empty pin.
The second particle
takes up the motion and again moves about a lattice constant.  Such 
pulse-like transport has been observed in a vortex system at drives
very close to depinning, and is imaged in Ref\cite{longaval}.
Again, it may be useful to think of the 
continuous motion of a
disturbance or soliton, rather than the jerky motion of individual particles,
as controlling the dynamics of the system.  How to identify this soliton  is
not as clear as in the case of the thermal system.  

At high drives, the lattice reorders and returns nearly to a perfect 
crystalline state, as illustrated in Fig.~\ref{fig:colloid2}.  The similarity
between decreasing the drive from very high values and increasing the
temperature in a clean system from low values was indicated by Koshelev
and Vinokur \cite{Koshelev}, although their theory holds only in the limit
of high drives when dislocations have not entered the lattice.
It would be very interesting to explore further the possible connections
between a lattice that has been disordered by temperature, and one that
has been disordered due to plastic flow.

\begin{figure}
  \begin{center}
    \begin{tabular}{c}
      \includegraphics[width=0.8\textwidth]{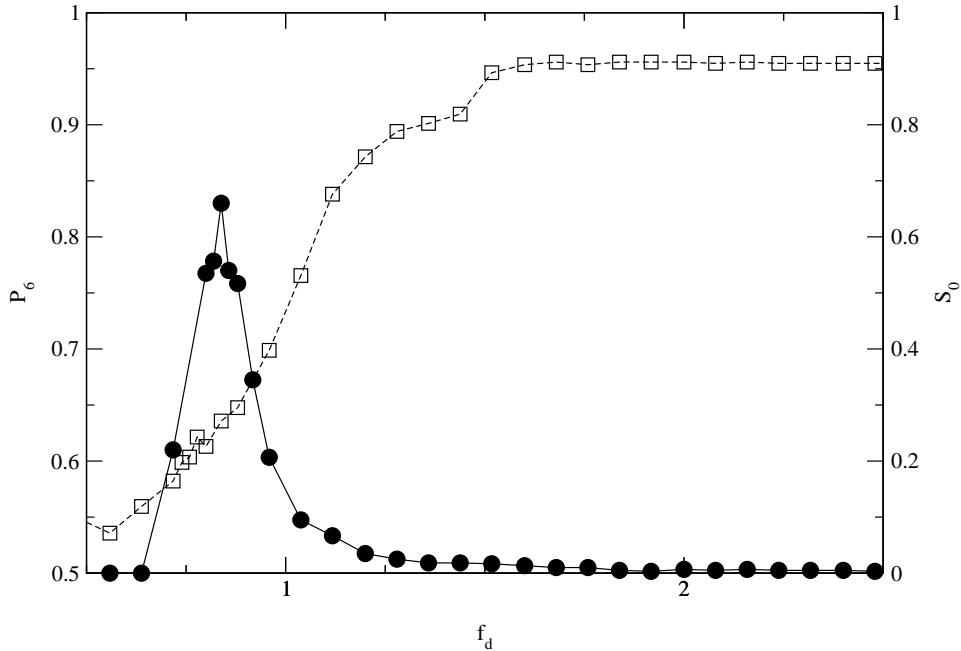}
    \end{tabular}
  \end{center}
  \caption[fig1]
  { \label{fig:volt}
Squares (dotted line): The fraction of sixfold coordinated particles $P_6$
versus driving force $f_d$, with reordering occurring at $f_d=1.6$.
Circles (solid line): The integrated noise power $S_0$, obtained from
$S(f)$, shows a peak in the plastic flow region.
}
\end{figure}

One notable similarity to the thermal case
can be found in the noise produced by a 
plastically flowing system.  Fig.~\ref{fig:volt} shows the fraction
of sixfold coordinated particles $P_6$ and the noise power $S_0$ obtained from
the time series of the particle velocity in the $x$ direction
as a function of driving force $f_d$,
both taken from a 2D simulation of vortices described
in Ref~\cite{voltnoise}.  
Here, the average vortex velocity
is  
the analog of an 
experimentally measured voltage signal. 
The
dynamic reordering transition at high drives to an elastically flowing
lattice occurs at a drive $f_d\approx 1.6$.  The noise power shows a peak in
the plastic flow regime, when the heterogeneity of the vortex motion
is at its maximum.  This peak is also associated with the greatest
transverse wandering of the vortices as they move, indicating that
the channel structure is highly braided \cite{river}.
The noise spectra is broad throughout the
plastic flow regime, with a $1/f^\alpha$ form.
This is quite similar to the behavior of the
noise power of the fluctuations in the dislocation number
in the thermal system considered in Section 3.
In both the plastic flow regime for the driven system, and the thermal
system just above melting, the dynamics are highly heterogeneous.
A deeper exploration of the similarities and differences in these
heterogeneous states may lead to new insights into both systems.

\section{Conclusion}

We have used noise measurements as a tool to explore the behavior
of both thermally melted and dynamically disordered systems in 1D
and 2D.  We find that above a threshold, which is a melting transition
in the thermal case, and a depinning transition in the driven case,
the dynamics of the system becomes highly heterogeneous.  In the
thermal system, domains of ordered particles are surrounded by
fluctuating grain boundaries, while in the driven system, regions of
pinned particles are surrounded by riverlike filaments of moving
particles.  In each case, the motion of the particles is not
continuous, but is characterized by long periods of no motion followed
by a rapid change in position by about a lattice constant.
In contrast, the motion of dislocations or solitons through the
system is much more continuous, indicating that the particle motion can
be considered as occurring whenever a dislocation goes past.
The power spectrum of a time series of 
the particle position, particle velocity, or
dislocation density has a $1/f^\alpha$ characteristic in this regime,
and the noise power peaks close to melting or depinning.  At higher
drives or lower temperatures, the system reorders and the noise
power drops.  These behaviors are not specific to a particular system,
as we have shown here by considering both superconducting vortices as
well as colloids in our simulations.  Our results suggest that there
may be interesting parallels between the heterogeneous nature of
the molten phase in the thermal system, and the plastically flowing
phase in the driven system.

\acknowledgments     
 
This work was supported by the US Department of Energy under
Contract No. W-7405-ENG-36.


\end{document}